\documentclass[a4paper,fleqn,usenatbib]{mnras}

\usepackage{mathptmx}

\usepackage[T1]{fontenc}
\usepackage{ae,aecompl}

\usepackage{graphicx}	
\usepackage{amsmath}	
\usepackage{amssymb}	
\usepackage{color}
\usepackage{textcomp}

\newcommand{\msun}{\mbox{M}_\odot}

\title[Long-lived structure in SDSS J1021+1744]{A large, long-lived 
structure near the trojan L5 point in the post common-envelope binary SDSS 
J1021+1744}

\author[P. Irawati et al.]{
P. Irawati$^{1}$,\thanks{E-mail: puji.irawati@narit.or.th}
A. Richichi$^{1}$,
M.~C.~P. Bours$^{2}$,
T.~R. Marsh$^{3}$,
N. Sanguansak$^{4}$, \newauthor
K. Chanthorn$^{4}$, 
J.~J. Hermes$^{2}$,
L.~K. Hardy$^{5}$, 
S.~G. Parsons$^{2}$, 
V.~S. Dhillon$^{5,6}$, \newauthor
S.~P. Littlefair$^{5}$ \\
\\
$^{1}$National Astronomical Research Institute of Thailand, 191 
Siriphanich Bldg., Huay Kaew Road, Chiang Mai 50200, Thailand\\ 
$^{2}$Departamento de F\'isica y Astronom\'ia, Universidad de 
Valpara\'iso, Avenida Gran Breta\~{n}a 1111, Valpara\'iso, Chile\\
$^{3}$Department of Physics, University of Warwick, Gibbet Hill Road, 
Coventry CV4 7AL, UK\\
$^{4}$School of Physics, Suranaree University of Technology,  
111 University Avenue, Muang, Nakhon Ratchasima 30000, Thailand\\
$^{5}$Department of Physics and Astronomy, University of Sheffield, 
Sheffield S3 7RH, UK\\
$^{6}$Instituto de Astrof\'isica de Canarias, 38205 La Laguna, Santa Cruz 
de Tenerife, Spain\\
}

\date{Accepted 2015 November 27. Received 2015 November 27; in original form 
2015 September 18}

\pubyear{2015}

\begin{document}
\label{firstpage}
\pagerange{\pageref{firstpage}--\pageref{lastpage}}
\maketitle

\begin{abstract}
SDSS~J1021+1744 is a detached, eclipsing white dwarf / M dwarf binary
discovered in the Sloan Digital Sky Survey. Outside the primary eclipse, 
the light curves of such systems are usually smooth and characterised by 
low-level variations caused by tidal distortion and heating of the M star 
component. 
Early data on SDSS~J1021+1744 obtained in June 2012 was unusual in 
showing a dip in flux of uncertain origin shortly after the white dwarf's 
eclipse. Here we present high-time resolution, multi-wavelength 
observations of 35 more eclipses over 1.3 years, 
showing that the dip has a lifetime extending over many orbits.
Moreover the ``dip" is in fact a series of dips 
that vary in depth, number and position, although they are always placed 
in the phase interval 1.06 to 1.26 after the white dwarf's eclipse, near 
the L5 point in this system. 
Since SDSS~J1021+1744 is a detached binary, it follows that the
dips are caused by the transit of the white dwarf by material
around the Lagrangian L5 point. A possible interpretation is that they are 
the signatures of prominences, a phenomenon already known from H$\alpha$ 
observations of rapidly rotating single stars as well as binaries. 
What makes SDSS~J1021+1744 peculiar is that the material is dense enough 
to block continuum light.
The dips appear to have finally faded out around 2015 May after the first 
detection by \citeauthor{2013MNRAS.429..256P} in 2012, suggesting a 
lifetime of years.
\end{abstract}

\begin{keywords}
binaries: close -- binaries: eclipsing -- stars: white dwarfs -- stars: 
individual: SDSS J102102.25+174439.9
\end{keywords}



\section{Introduction}

In recent years, primarily as the result of the Sloan Digital Sky Survey
(SDSS, \citealt{2000AJ....120.1579Y,2009ApJS..182..543A}), large numbers 
of white dwarf / main-sequence (WDMS) binaries have been discovered 
\citep{2007MNRAS.382.1377R,2012MNRAS.419..806R,2013MNRAS.433.3398R}. 
A significant number of these have periods so short that they must have 
emerged from a phase in which both stars orbited within the envelope of 
the white dwarf's progenitor. 
During such ``common envelope'' phases, binary orbital energy is lost to 
the envelope \citep{1984ApJ...277..355W}, resulting in the observed 
short-periods 
(or often complete merging of the stars, \citealt{2015MNRAS.447.1713B}). 
White dwarf / main-sequence post common-envelope binaries (PCEBs) form a 
large, easily observed population for testing the outcome of the 
common-envelope phase, which is significant in the formation of many 
classes of close binary.

As the number of known PCEBs has increased, so too has the number of 
eclipsing systems. Thus, while in 2000 we knew of just 5 eclipsing PCEBs
\citep[][including the WD + K dwarf system V471~Tau]{2000NewAR..44..119M}, 
the most recent census \citep{2015MNRAS.449.2194P} lists 71 such systems. 
Amongst these is the subject of the present paper, 
SDSS~J102102.25+174439.9 (hereafter J1021+1744). 
J1021+1744 was first 
recognized as a WDMS binary from the SDSS Data Release 7 and the stellar 
parameters of this binary were published as part of the online 
SDSS WDMS binary catalogue (http://www.sdss-wdms.org, 
\citealt{2012MNRAS.419..806R}). This binary was also suspected as strong 
candidate PCEB from its radial velocity variability. The catalogue 
published an M4 type for the red dwarf star and a white dwarf with mass of 
$1.06 \pm 0.087 \,\msun$. The effective temperature of the white dwarf 
in J1021+1744 was given as `hot' and `cold' solution from the 
Balmer line profile fits. The white dwarf temperatures are $32595 \pm 928$ 
K and $17505 \pm 820$ K for the hot and cold case, respectively.

The eclipsing nature of J1021+1744 was discovered by 
\citet[P13 hereafter]{2013MNRAS.429..256P} from a search for photometric 
variability of WDMS systems in Catalina Sky Survey (CSS, 
\citealt{2009ApJ...696..870D,2014MNRAS.441.1186D}) data. 
The eclipses are of the white dwarf by its M dwarf companion and recur 
with an orbital period of $0.14\,$ days.  Using the robotic Liverpool 
Telescope (LT) in 2012 June, P13 found a drop in 
the brightness shortly after eclipse, about half as deep as 
the eclipse itself. 
This is highly unusual: outside eclipse, the vast majority of these 
systems show only slow variations due to irradiation and tidal distortion. 
P13 showed a possible flare taking place 
before the white dwarf was fully out of the eclipse, leading them to 
suggest that the dip in flux might be caused by material ejected from the 
flare. They gave new constraints for the white dwarf mass using their new 
ephemeris data, the measured radial velocity, and the mass function 
equation, lowering the estimated mass to $0.50\pm0.05 \,\msun$. 
\citet{2012MNRAS.419..806R}'s white dwarf mass was based upon model 
atmosphere fitting, made difficult because of the contamination 
of the white dwarf's spectrum by its companion. 

In this paper we present photometric observations of J1021+1744 taken 
mainly with the 2.4\,m telescope at the Thai National Observatory, 
covering more than 30 eclipses from 2014 January to 2015 May in a variety 
of filters and with sub-minute time resolution. Our observations reveal 
that the dip observed by P13 is long-lived, with a lifetime of at least a 
few years. We also show that the dip is resolved into multiple components 
that vary both with time and wavelength. We suggest here that the dips 
originate from obscuration of the white dwarf showing that this detached 
binary is able to support dense clouds of material around the L5 trojan 
point. 

\section[]{Observations And Data Reduction}

The bulk of our photometric data of J1021+1744 were taken using the  
2.4\,m Thai National Telescope (TNT) on Doi Inthanon, 
equipped with the ULTRASPEC camera. This facility is ideal for such 
studies, thanks to the combination of high time resolution, sensitivity 
and flexibility in time allocation. We supplemented this with a single 
eclipse observed with the high-speed triple-beam camera ULTRACAM mounted 
on the 4.2m William Herschel Telescope (WHT) on 2015 January 17.
ULTRASPEC is based on a low-noise 1k $\times$ 1k EMCCD frame-transfer 
detector, and is described in detail by \citet{2014MNRAS.444.4009D}. 
During the first observing cycle of TNT (2013 November -- 2014 April), we 
monitored this star for 17 nights from 2014 January to April, covering 
more than 20 eclipses in different filters. In the following cycle 
(2014 November -- 2015 May), we obtained 13 more eclipses from 9 nights 
of observations. The log of our observations is presented in 
Table \ref{tab1}.

\begin{table*}
 \centering
  \caption{Observation log of J1021+1744 obtained from TNT using 
  ULTRASPEC and WHT with ULTRACAM. Each row represents one dataset 
  where the start and the end of the exposure are given in columns 4 
  and 5, together with the corresponding orbital phase in column 6. 
  The filters are listed in column 7. The presence of a colon before
  the filter denotes that the nominal filter name has been changed
  to an adopted filter, as explained in the text. Column 8 gives 
  the exposure time of a single frame, where there is a dead time of 
  14.9\,ms between exposures. In Column 9 and 10 we list SNR (measured in 
  the pre-eclipse part; see text) and approximate seeing values for each 
  run.}
  \label{tab1}
  \begin{tabular}{cccccccccc}
  \hline
  No & Date & Telescope & UT start & UT end & Orbital Phase & Filter & 
  Sampling & SNR & Seeing \\ 
    &  &  & & & coverage & & (sec) & & ($''$) \\
  \hline
  1 & 2014 Jan 07 & TNT & 20:27:49 & 22:10:17 & 0.81 -- 1.32 & $g'$ & 
  59.670 & 37 & $2.0-3.0$ \\
  2 & 2014 Jan 08& TNT & 16:56:17 & 18:14:52 & 0.88 -- 1.28 & :clear & 
  59.670 & 132 & $1.0-2.0$ \\
  3 & 2014 Jan 08 & TNT & 20:02:05 & 21:23:39 & 0.80 -- 1.21 & :clear & 
  59.670 & 153 & $1.2-1.5$ \\
  4 & 2014 Jan 10& TNT & 16:01:33 & 17:13:00 & 0.86 -- 1.22 & $r'$ & 
  55.667 & 38 & $1.8-3.2$ \\
  5 & 2014 Jan 11 & TNT & 15:47:19 & 16:36:49 & 0.92 -- 1.17 & :$z'$ & 
  43.667 & 68 &  $1.2-1.5$ \\
  6 & 2014 Jan 11 & TNT & 18:58:30 & 20:16:02 & 0.87 -- 1.25 & :$i'+z'$ & 
  24.777 & 122 & $0.9-1.1$\\
  7 & 2014 Jan 11 & TNT & 21:53:02 & 23:23:08 & 0.73 -- 1.17 & $r'$ & 
  34.777 & 71 & $1.4-1.8$ \\
  8 & 2014 Jan 12 & TNT & 18:14:40 & 19:45:29 & 0.77 -- 1.22 & :$i'+z'$ & 
  12.772 & 91 & $1.0-1.8$ \\
  9 & 2014 Jan 12 & TNT & 22:03:15 & 23:10:48 & 0.90 -- 1.24 & :KG5 & 
  49.777 & 85 & $1.2-1.4$ \\
 10 & 2014 Jan 15& TNT & 20:21:41 & 22:22:30 & 0.78 -- 1.38 & :$r'$ & 
 58.777 & 17 & $2.0-3.0$ \\
 11 & 2014 Jan 28 & TNT & 15:04:48 & 16:37:17 & 0.83 -- 1.29 & KG5 &  
 9.852 & 31 & $1.4-2.0$ \\
 12 & 2014 Jan 31 & TNT & 16:54:42 & 20:29:19 & 0.75 -- 1.81 & $i'$ & 
 3.352 & 38 & $1.2-3.5$ \\
 13 & 2014 Feb 11 & TNT & 19:35:39 & 21:00:47 & 0.98 -- 1.34 & $g'$ & 
 12.777 & 19 & $1.6-2.1$\\
 14 & 2014 Feb 28 & TNT & 18:38:05 & 20:31:30 & 0.75 -- 1.32 & $g'$ & 
 3.352 & 21 & $1.2-2.0$ \\
 15 & 2014 Mar 26 & TNT & 14:56:13 & 16:47:43 & 0.89 -- 1.44 & $g'$ & 
 9.872 & 27 & $1.4-1.8$ \\
 16 & 2014 Mar 29 & TNT & 13:10:23 & 15:11:32 & 0.74 -- 1.34 & $r'$ & 
 9.852 & 26 & $1.2-2.2$ \\
 17 & 2014 Mar 30 & TNT & 12:50:12 & 15:02:53 & 0.76 -- 1.42 & $g'$ & 
 3.352 & 20 & $1.2-2.0$ \\
 18 & 2014 Mar 31 & TNT & 12:32:27 & 17:26:53 & 0.80 -- 2.26 & $r'$ & 
 3.352 & 24 & $1.2-2.2$ \\
 19 & 2014 Apr 01 & TNT & 12:23:07 & 17:09:11 & 0.88 -- 2.29 & $g'$ & 
 3.352 & 18 & $1.3-2.0$ \\
 20 & 2014 Apr 02& TNT & 14:04:14 & 18:14:38 & 0.50 -- 1.74 & :$g'$ & 
 4.852 & 15 & $1.2-1.8$ \\
 21 & 2014 Apr 03& TNT & 16:58:35 & 19:32:28 & 0.49 -- 1.25 & $r'$ & 
 3.352 & 16 & $1.0-2.2$ \\
 22 & 2014 Dec 22 & TNT & 17:09:37 & 20:29:12 & 0.30 -- 1.29 & $g'$ & 
 4.852 & 22 & $1.8-3.0$ \\
 23 & 2015 Jan 01 & TNT & 20:44:18 & 23:06:57 & 0.92 -- 1.32 & KG5 & 
 10.352 & 52 & $1.2-1.5$ \\
 24 & 2015 Jan 12 & TNT & 20:32:47 & 21:33:49 & 0.93 -- 1.24 & $g'$ & 
 9.876 & 17 & $1.6-3.0$ \\
 25 & 2015 Jan 17 & WHT & 01:12:31 & 02:55:59 & 0.82 -- 1.33 & $u'g'r'$ & 
 12;4;4 & 48 & $1.0-2.0$ \\
 26 & 2015 Feb 19 & TNT & 16:44:53 & 19:28:52 & 0.55 -- 1.36 & $g'$ & 
 9.352 & 25 & $1.0-2.2$ \\
 27 & 2015 Feb 19& TNT & 19:31:20 & 22:52:28 & 0.38 -- 1.37 & $r'$ & 
 9.352 & 38 & $1.0-2.0$ \\
 28 & 2015 Feb 24 & TNT & 15:47:19 & 17:08:34 & 0.89 -- 1.29 & KG5 &  
 9.862 & 42 & $1.5-2.2$ \\
 29 & 2015 Mar 18 & TNT & 12:57:49 & 16:06:24 & 0.79 -- 1.72 & $r'$ & 
 9.352 & 14 & $1.2-2.2$ \\
 30 & 2015 Mar 18 & TNT & 16:09:00 & 19:16:25 & 0.74 -- 1.66 & $g'$ & 
 9.352 & 30 & $1.2-2.2$ \\
 31 & 2015 Mar 19 & TNT & 12:28:51 & 14:11:28 & 0.77 -- 1.28 & $i'$ & 
 6.000 & 38 & $1.0-1.6$ \\
 32 & 2015 Mar 19 & TNT & 15:38:22 & 17:44:46 & 0.71 -- 1.33 & $g'$ & 
 9.352 & 29 & $1.3-1.8$ \\
 33 & 2015 Mar 19 & TNT & 18:47:17 & 20:51:36 & 0.64 -- 1.26 & KG5 & 
 7.352 & 32 & $1.6-3.0$ \\
 34 & 2015 Mar 20 & TNT & 12:16:12 & 13:36:15 & 0.83 -- 1.23 & $g'$ & 
 9.352 & 8 & $1.6-4.0$ \\
 35 & 2015 May 12 & TNT & 12:45:53 & 15:03:51 & 0.90 -- 1.24 & $g'$ & 
 9.352 & 6 & $1.9-2.6$ \\ 
  \hline
\end{tabular}
\end{table*}

Each observation consists of several hundreds to several thousands of 
frames, with the sampling times listed in Table~\ref{tab1}. The detector 
integration times are 14.9 ms shorter than the sampling times 
(see section 3.4 of \citealt{2014MNRAS.444.4009D}). 
The frame size is usually equal to the full detector window, 
although in some cases smaller windows are adopted (e.g. 400$\times$800 
pixels). Each frame is accurately time-stamped at mid-exposure thanks to 
a dedicated GPS system. The data are then processed using the 
ULTRACAM pipeline \citep{2007MNRAS.378..825D}. After bias subtraction and 
flat fielding, the data are corrected to take into account the position 
of the Solar System Barycenter. One or more reference stars are 
recorded simultaneously with J1021+1744, allowing us to obtain accurate 
relative photometry. We note that the ULTRACAM pipeline includes adaptive 
estimates of seeing and star positions, and is thus very robust against 
changes in photometric quality, airmass, and tracking errors.

For our observations we used several ULTRASPEC filters which are
similar to the passbands of the SDSS photometric system, namely
$g', r', i', z'$. Additionally, we used the KG5 filter which
is effectively equivalent to SDSS $u'+g'+r'$ (as described in 
\citealt{2014MNRAS.444.4009D}), and the self-explanatory $i'+z'$ filter. 
Finally, we also obtained data with a clear filter (white light).  

During the data processing we noticed several inconsistencies in the 
depth of the primary eclipses. Eclipses in white dwarf binaries are 
wavelength dependent, and we realized that some of our light curves 
supposedly taken with the same filter appear to have different eclipse 
depths. Further investigations indicated that there were problems 
with the filter wheel rotation on some of our nights. In other words, 
the wheel did not move to the intended filter, without any alerts at the 
software level. We recovered from the problem as follows.

For each of the data sets with an ambiguous filters, we stacked all frames 
to create one deep image and examined all non-variable stars in the field. 
We then compared the fluxes against various sky surveys. Using this 
method we could identify the correct filter for all of our affected data.
We also verified the eclipse depths by filter, as discussed
in Section~\ref{sec:Photom}.
The changes from nominal to adopted filters are marked in  
Table~\ref{tab1}. 
The problem was fixed at the hardware level in the summer of 2014 and is 
not present in later data.

In Table~\ref{tab1} we list the signal-to-noise ratio (SNR) as computed 
over 10 minutes in the pre-eclipse part of the light curve, centered 
around phase 0.93. The numbers in the seeing column are approximate 
values measured from the stellar profiles.


\section[]{Photometric analysis}\label{sec:Photom}

\subsection{Light curve fitting}\label{subsec:LCfit}

We implemented a light curve fitting method \citep{2010MNRAS.402.1824C} to
obtain the parameters for our binary model. 
To find the best parameters of the model, we first fit all 
light curves with the same filter using an initial model. In this model, 
we allowed the inclination angle, the white dwarf and red dwarf radii, 
and the red dwarf temperature to vary, while the mass ratio 
and the white dwarf temperature are fixed. We have excluded those parts of 
light curves with dips and other variations during the fitting process.
The `hot' solution with $32595$ K for the temperature of the white 
dwarf gives too strong a reflection effect in our model. On the other 
hand, the lower temperature of $17505$ K from the SDSS WDMS binary 
catalogue is probably also not reliable (it has a very high gravity of 
$\log (g) = 9.5$ and it is found at the edge of the model grid), but 
must be closer to the correct value. Hence, we chose the value of $17505$ 
K for our model. The strong contamination by the red dwarf and the 
faintness of the system are possibly the cause of the uncertainty in the 
temperature determination.  

We also applied a Markov Chain Monte Carlo (MCMC) algorithm 
to confirm the result of the light curve fit. 
Using a fixed mass ratio of $q=0.5$ with white dwarf temperature of 
$T_\mathrm{WD}=17505$ K, our best fit model gives an 
inclination angle of $i=85^{\circ}$. The radii of the two stars 
(scaled by the binary separation) are $R_\mathrm{WD}/a=0.0116$ 
and $R_\mathrm{sec}/a=0.3572$, with the red dwarf companion almost 
filling its Roche Lobe. The temperatures of the red dwarf star derived 
from our model is $T_\mathrm{sec}=3160$ K. 
We then fitted each individual light curve using the 
binary parameters given above, allowing only the orbital period and the 
time of mid-eclipse as free parameters.

\subsection{Mid-eclipse times and new ephemeris}
\label{subsec:ephemeris}

\begin{table*}
 \centering
  \caption{The eclipse times for J1021+1744. For each date, we listed the 
  filter names, cycle number, mid-eclipses, the O$-$C and the 
  uncertainties in our O$-$C calculation in seconds. The eclipse times of 
  the original and the new LT data are given in the first and second rows. 
  For ULTRACAM data (2015 Jan 17), we list the weighted average of the 
  mid-eclipse times from each filter.}
  \label{tab2}
  \begin{tabular}{|c|c|c|l|r|@{}c@{}|r|}
  \hline    
  Date & Filter & Cycle  & Mid-eclipse time & \multicolumn{3}{c}{O$-$C} \\
        &        & number & BMJD(TDB) & \multicolumn{3}{c}{(sec)} \\
  \hline 
 2012 Jun 15 & V$+$R & -4068 & 56093.90558(12) & 57.99 & $\pm$ & 10.56 \\
             &       &       & 56093.905144(79) & 20.34 & $\pm$ & 7.10 \\
 2014 Jan 07 & $g'$  & 0 & 56664.884326(23) & 0 & $\pm$ & 2.82 \\
 2014 Jan 08 & clear & 6 & 56665.726518(49) & 3.40 & $\pm$ & 4.67 \\
 2014 Jan 08 & clear & 7 & 56665.866817(46) & -1.74 & $\pm$ & 4.42 \\
 2014 Jan 10 & $r'$ & 20 & 56667.691512(61) & 0.91 & $\pm$ & 5.63 \\
 2014 Jan 11 & $z'$ & 27 & 56668.67420(13) & 15.96 & $\pm$ & 11.26 \\
 2014 Jan 11 & $i'+z'$ & 28 & 56668.814482(79) & 9.60 & $\pm$ & 7.13 \\
 2014 Jan 11 & $r'$ & 29 & 56668.954726(50) & -0.37 & $\pm$ & 4.76 \\
 2014 Jan 12 & $i'+z'$ & 35 & 56669.797077(56) & 16.80 & $\pm$ & 5.25 \\
 2014 Jan 12 & KG5 & 36 & 56669.937231(17) & -0.88 & $\pm$ & 2.49 \\
 2014 Jan 15 & $r'$ & 57 & 56672.884782(93) & 0.57 & $\pm$ & 8.26 \\
 2014 Jan 28 & KG5 & 148 & 56685.657432(21) & 0.87 & $\pm$ & 2.72 \\
 2014 Jan 31 & $i'$ & 170 & 56688.745118(33) & 25.67 & $\pm$ & 3.47 \\
 2014 Feb 11 & $g'$ & 249 & 56699.833578(47) & -6.79 & $\pm$ & 4.56 \\
 2014 Feb 28 & $g'$ & 370 & 56716.817073(14) & 0.60 & $\pm$ & 2.31 \\
 2014 Mar 26 & $g'$ & 554 & 56742.643108(20) & 2.70 & $\pm$ & 2.65 \\
 2014 Mar 29 & $r'$ & 575 & 56745.590612(30) & 0.09 & $\pm$ & 3.25 \\
 2014 Mar 30 & $g'$ & 582 & 56746.573142(14) & 1.72 & $\pm$ & 2.33 \\
 2014 Mar 31 & $r'$ & 589 & 56747.555624(31) & -0.81 & $\pm$ & 3.31 \\
 2014 Mar 31 & $r'$ & 590 & 56747.696023(27) & 2.70 & $\pm$ & 3.09 \\
 2014 Apr 01 & $g'$ & 596 & 56748.538134(22) & -0.94 & $\pm$ & 2.74 \\
 2014 Apr 01 & $g'$ & 597 & 56748.678507(16) & 0.33 & $\pm$ & 2.41 \\
 2014 Apr 02 & $g'$ & 604 & 56749.660988(22) & -2.25 & $\pm$ & 2.73 \\
 2014 Apr 03 & $r'$ & 612 & 56750.783865(43) & -1.71 & $\pm$ & 4.25 \\
 2014 Dec 22 & $g'$ & 2486 & 57013.816200(15) & 0.67 & $\pm$ & 2.40 \\
 2015 Jan 01  & KG5 & 2558 & 57023.922030(12) & 0.68 & $\pm$ & 2.25 \\
 2015 Jan 12  & $g'$ & 2636 & 57034.870024(22) & 1.59 & $\pm$ & 2.74 \\
 2015 Jan 17 & $u'g'r'$ & 2666 & 57039.0808021(41) & 2.28 & $\pm$ & 2.02 \\
 2015 Feb 19 & $g'$ & 2906 & 57072.766864(12) & -0.45 & $\pm$ & 2.23 \\
 2015 Feb 19 & $r'$ & 2907 & 57072.907249(20) & 1.77 & $\pm$ & 2.66 \\
 2015 Feb 24 & KG5 & 2941 & 57077.679420(20) & -0.49 & $\pm$ & 2.67 \\
 2015 Mar 18 & $r'$ & 3097 & 57099.575392(27) & -1.38 & $\pm$ & 2.34 \\
 2015 Mar 18 & $g'$ & 3098 & 57099.715735(14) & -0.031 & $\pm$ & 3.75 \\
 2015 Mar 19 & $i'$ & 3104 & 57100.557875(56) & 0.25 & $\pm$ & 2.38 \\
 2015 Mar 19 & $g'$ & 3105 & 57100.698265(15) & -2.45 & $\pm$ & 5.26 \\
 2015 Mar 19 & KG5 & 3106 & 57100.838629(24) & 0.73 & $\pm$ & 2.87 \\
 2015 Mar 20 & $g'$ & 3111 & 57101.540570(50) & 13.42 & $\pm$ & 4.73 \\
 2015 May 12 & $g'$ & 3489 & 57154.596013(21) & -0.95 & $\pm$ & 2.71 \\
  \hline
\end{tabular}
\end{table*}

We first adopted the ephemeris from P13 to compute the orbital phase, 
where the orbital period is $P_\mathrm{orb}$=0.140359073(1) days. 
The adopted ephemeris shows that the mid-eclipse is offset earlier 
by $\sim$3~min from the expected time. The derived O$-$C values 
from ULTRASPEC data taken in late 2014 and 2015 showed a linear but 
decreasing trend, indicating that the adopted P13's ephemeris is not 
suitable for our data. We calculated a new ephemeris for J1021+1744 
where we used a light curve fitting method (as described above) to 
find the mid-eclipse timings for every light curve.
The new orbital period resulting from our fitting process is shorter 
by almost 0.03\,s and the new ephemeris derived from our data is 
\begin{equation*}
\mathrm{BMJD(TDB)} = 56664.8843262(231) + 0.140358755(1) E 
\end{equation*}

\begin{figure*}
 \centering
 \includegraphics[scale=0.7]{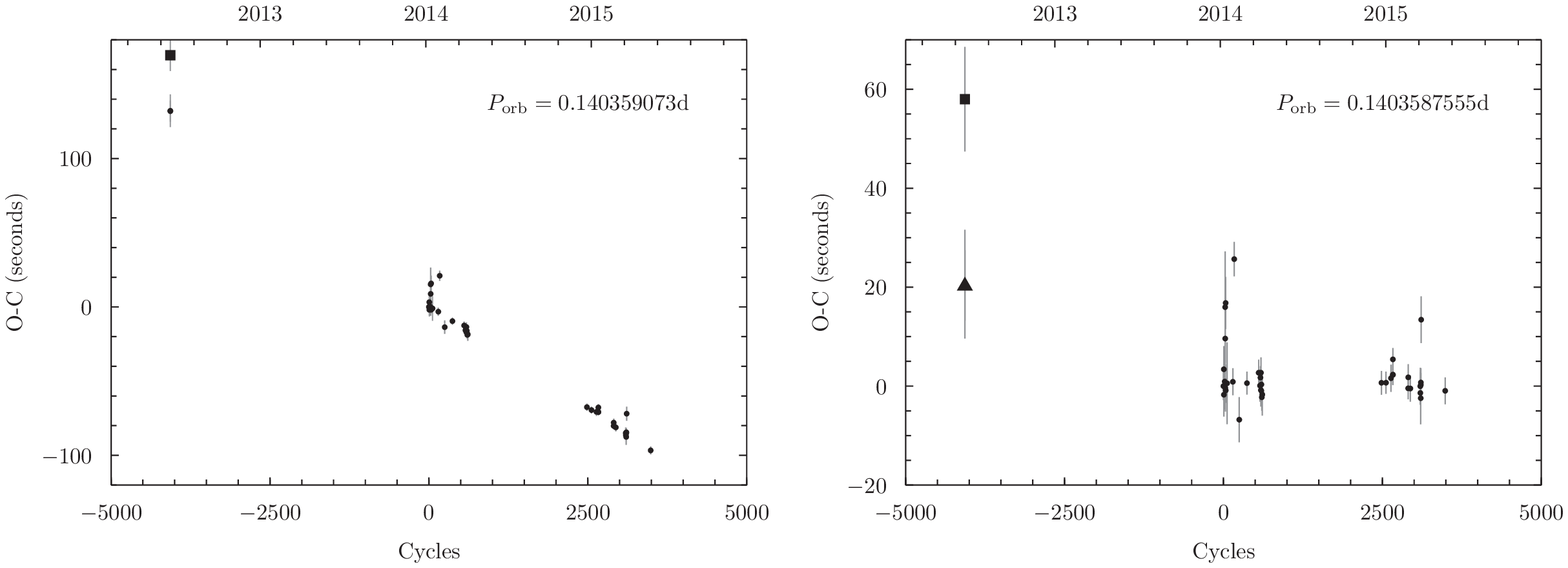}
 \caption{O$-$C diagrams for J1021+1744 calculated using the old 
 ephemeris of P13 (left) and our new ephemeris 
 from ULTRASPEC data (right). The LT data from P13 
 is marked with a filled square. Our re-fitting result to the LT data 
 using the  ephemeris from ULTRASPEC data is marked with a filled triangle 
 (see 
 section \ref{subsec:ephemeris}).}
 \label{ocdiagram}
\end{figure*}

We list the mid-eclipse times in Table~\ref{tab2}, including the 
mid-eclipse of the LT light curve of P13. We have applied a barycentric 
correction to all of our times following a method developed by 
\citet{2010PASP..122..935E}, and we present these numbers in BMJD(TDB). 
The O$-$C are derived with respect to the T0 on 2014 January 7 (the date 
of the first ULTRASPEC data obtained at TNT). 
In Figure~\ref{ocdiagram} we compare the O$-$C values calculated 
using P13's orbital period (left panel) with the values from our newly 
derived orbital period (right panel). P13's LT data point is plotted as a 
filled square. Additionally, we fitted the LT light curve using our binary 
model and recalculated the O$-$C using our ephemeris (presented as filled 
triangle). 
There is a 38 seconds difference between the original and the new LT 
mideclipse times. In their paper, P13 mentioned a flare which occured 
during the egress of the white dwarf (see Figure 5 of P13). 
This flare could have affected the fitting of the eclipse in P13. Since we 
know the width of the eclipse from our ULTRASPEC data, we can exclude the 
flare in P13 data for our light curve fitting.

\subsection{Dips in J1021+1744}

\begin{figure}
 \centering
 \includegraphics[scale=0.73]{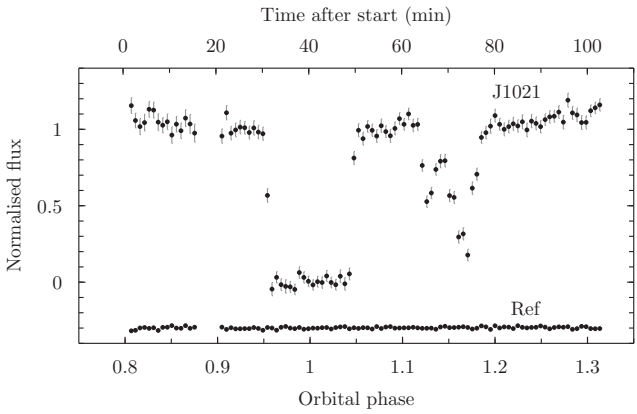}
 \caption{$g'$ filter light curve of J1021+1744 from 2014 January 7, 
 displaying the white dwarf eclipse and double dip from orbital 
 phase 1.1 to 1.2. Each data point represents a single frame with 
 exposure time 59~sec. The small gap around phase 0.89 is due to the 
 rotation of the instrument during the exposure.}
 \label{7jan}
\end{figure}

We have detected for the first time clear evidence of multiple dips 
after the main eclipse in the light curve of J1021+1744, as shown in 
Figure~\ref{7jan}. The light curves presented in the figure are the 
$g'$ filter data of our target and the comparison star (marked as ``Ref") 
taken on the night of 2014 January 7. In this work, we present the light 
curves in terms of orbital phase, using the ephemeris derived from 
ULTRASPEC data. For reference, a 0.1 phase interval corresponds to 
$\sim$20 minutes. 

The flux scale is normalised and then rescaled to 1 outside the eclipse 
and to 0 during the eclipse. We computed the average values between the 
orbital phase $0.9-0.95$ and $0.97-1.03$, and used the first phase range 
for our light curve normalisation. The rescaling factor is the difference 
between the two average values. We chose to rescale our light curves to 
minimise the effect of the filters over our analysis of the dips (see 
below).  

We used a similar approach for the light curve of the comparison star 
shown in Figure~\ref{7jan}. We normalised the flux of the comparison star 
to 1 using the average value in the same phase range mentioned above. 
The normalized comparison star's flux is then offset to -0.3 for clarity. 

The light curve of J1021+1744 clearly shows variations after the white 
dwarf eclipse, around phase $1.1-1.2$. The drop in our light curve has two 
connected dips with different depth. The system dims by $\sim$50\% during 
the first dip, and becomes even fainter through the second dip, with more 
than 80\% of the light of the white dwarf blocked. The star then slowly 
returns to its out-of-eclipse brightness around phase 1.2. 
The first dip seems to be consistent in phase and in amplitude with 
that reported by P13. The second dip however was not present at all in the 
LT light curve. It should have been clearly evident as it is wider 
and deeper in our data than the first dip. 
We suspect that the second dip developed after the P13 observation 
in 2012 June. We will show later that these dips are evolving in shape and 
amplitude. 

\begin{figure}
 \centering
 \includegraphics[scale=0.73]{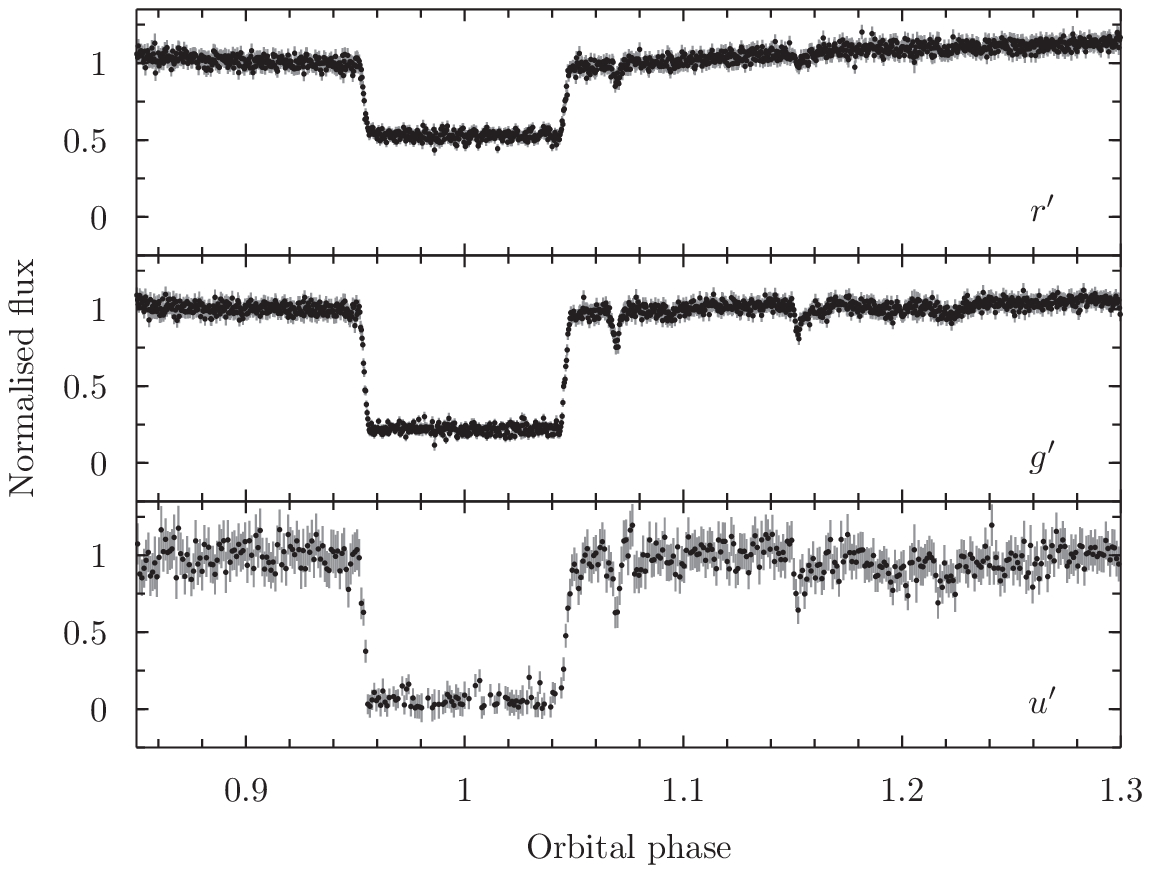}
 \caption{Light curve of J1021+1744 taken on 2015 January 17 from the WHT 
 telescope with ULTRACAM. From top to bottom the filters are $r'$, $g'$, 
 and $u'$. The data in $u'$ has fewer points due to longer 
 exposure time. The light curves are normalised to the average flux 
 between phase 0.9$-$0.95. The white dwarf eclipses follow the expected 
 pattern where they are deeper at bluer wavelengths. The dip features 
 seem to follow a similar trend.}
 \label{wdeclipse}
\end{figure}

After the initial work to identify the correct filter, the 
wavelength-dependence of the primary eclipses of J1021+1744 follow our 
expectations for this type of binaries. 
The eclipses are dominant in the blue and decrease in depth towards the 
red part of the spectrum. The trend in wavelength for the dip features is 
the same as in the main eclipse, and possibly even more pronounced.
The dips are obscuration of the white dwarf, hence the similarity 
in the wavelength dependency between the eclipses and the dips. 
Later data taken (simultaneously) using ULTRACAM show that the dips are 
deepest in the $u'$ band, as illustrated in Figure~\ref{wdeclipse}. 
The ULTRACAM $u'$ band data were taken with a longer exposure time 
compared to the $g'$ or $r'$ filters, therefore we have smaller number of 
datapoints in the $u'$ band light curve. For these data, the sampling in 
the $u'$ filter is three times slower than in the other filters. We 
normalised the flux scale to the average value of the pre-eclipse section 
only (phase $0.90-0.95$) to show the relative shape and the depth of the 
white dwarf eclipses. We would like to note that the ULTRACAM data 
definitely confirms the dip depths as a function of wavelength, as the 
ULTRASPEC data were taken in different orbits for different filters.

The dips are clearly seen in the $g'$ filter, though they appear to be 
less prominent at this wavelength. They are shallower in $r'$ and KG5 
filters, and barely visible in the $i'$ and $i'+z'$ filters. This may 
suggest that the material causing these dips has an optical thickness 
decreasing towards the red part of the spectrum.  

\begin{figure*}
 \centering
 \includegraphics[scale=1]{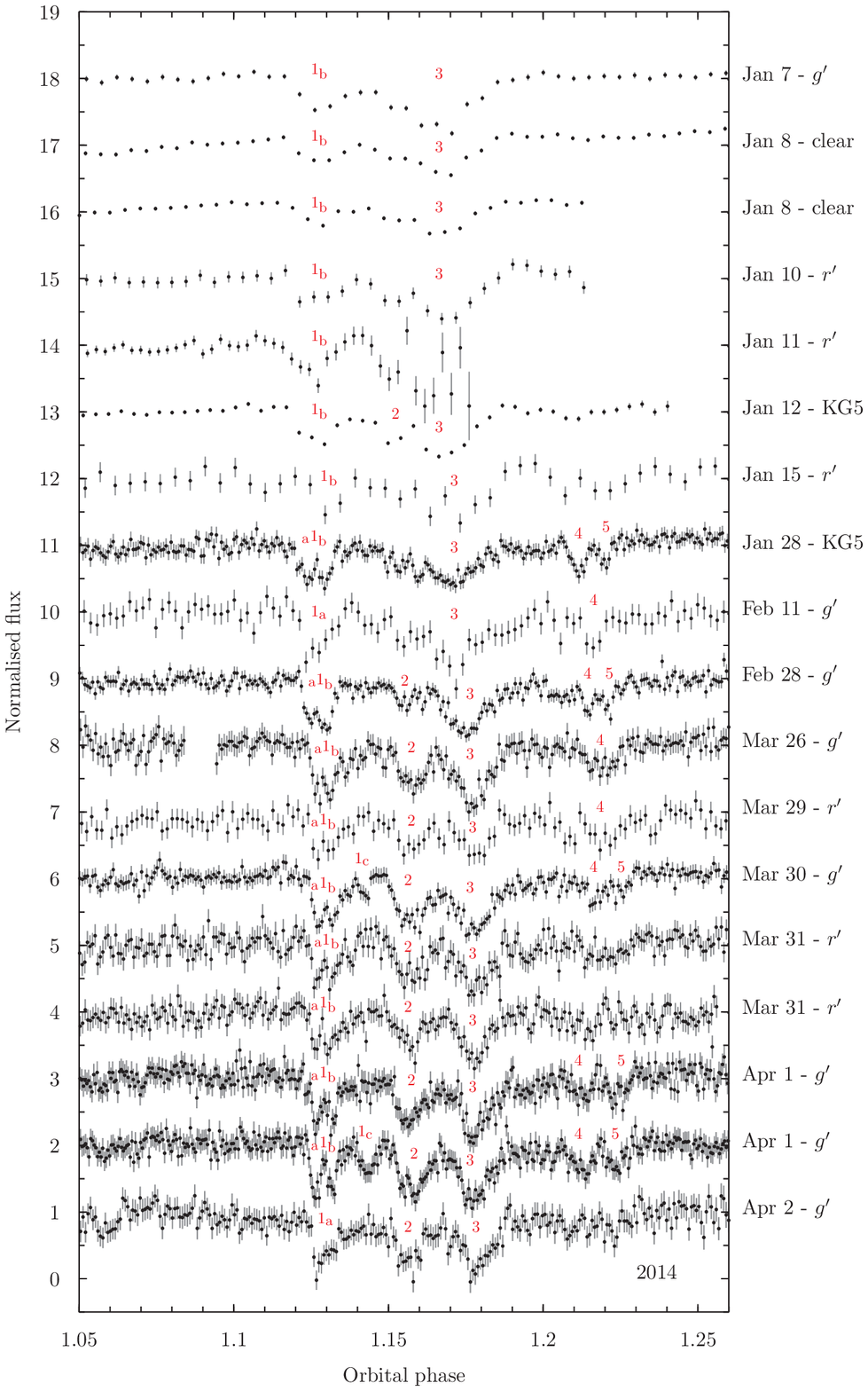}
 \caption{A close up look of the light curves of J1021+1744 (2014 data) 
 between phases $1.05-1.26$. The light curves are arranged from the oldest 
 at the top to the newest at the bottom. Some data with short exposure 
 times are binned for clarity. The dips in our 2014 light curves 
 clearly evolves in shape, width, and depth. We mark the position 
 of each dip with numbers from 1 to 5. In some light curves where dip 1
 is split into three smaller dips, we annotated them with $1_{\small 
  \textrm a}$, $1_{\small \textrm b}$ and $1_{\small \textrm c}$. 
 Each mark represents one dip, except for ${}_{\small \textrm a}1_{\small 
 \textrm b}$ where we count two dips.
 The total number of dips for each light curve is counted based on 
 these marks.}
 \label{dips2014}
\end{figure*}  

\begin{figure*}
 \centering
 \includegraphics[scale=1.1]{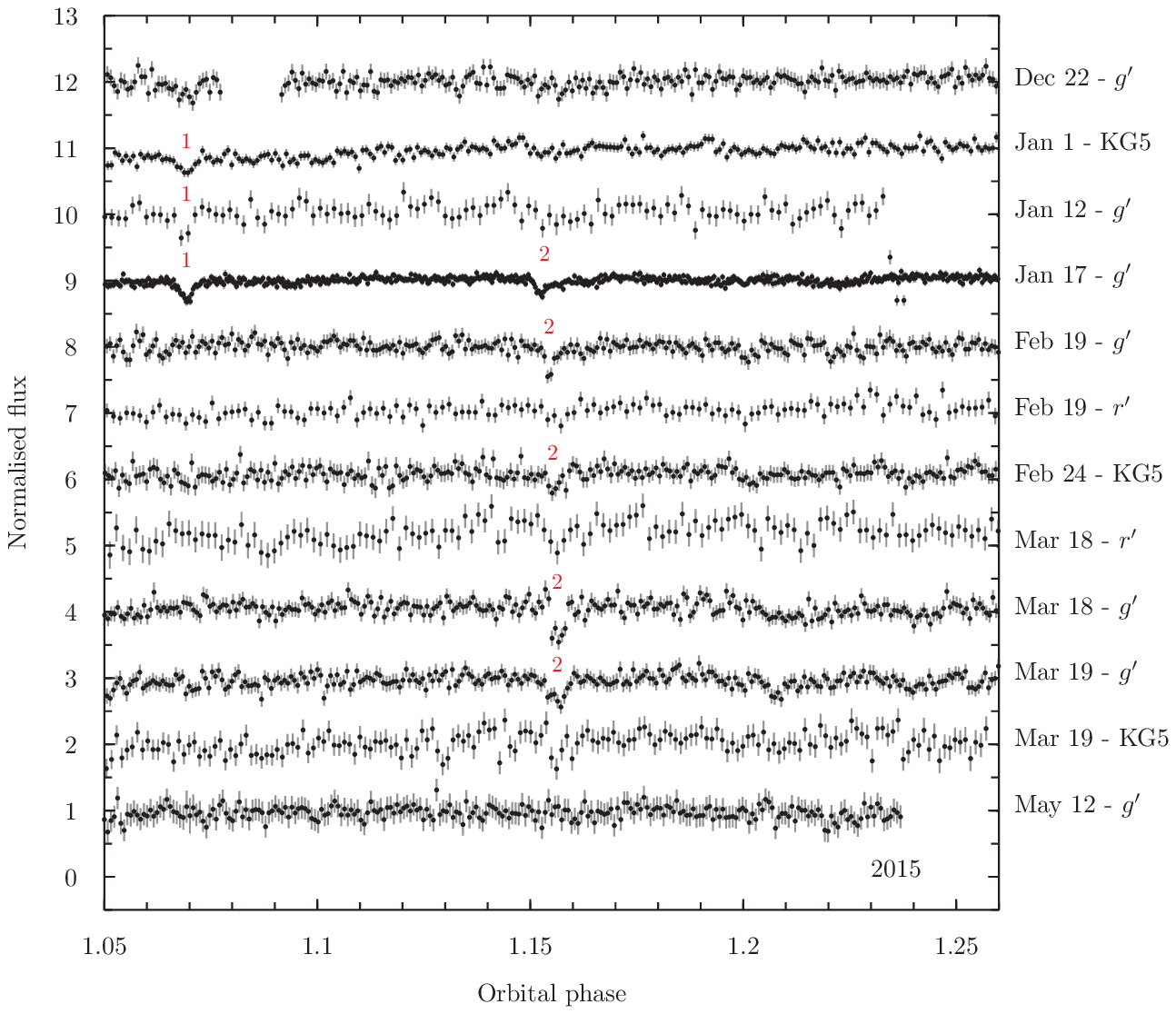}
 \caption{Similar to Figure 4, but for the data obtained in 2015 observing 
 season. The first data were taken on 2014 December 22. Only one or two 
 small dips are seen in some light curves. The light curves with the 
 highest time-resolution have been binned for clarity.}
 \label{dips2015}
\end{figure*}

Our next task was to examine the dips profile. Since 
the mid-eclipse times are known, the location of the dips can be 
determined accurately. In our 2014 data, the dips were prominent and were  
always located between phase 1.10 and 1.25. Multiple number of dips are 
recorded in every light curve, often with complex shapes. 
We present the light curves obtained between 2014 January$-$April in 
Figure~\ref{dips2014}, focusing on the section where the dips are visible.
The light curves are ordered in time from top to bottom. 
We exclude the data in the $i'$, $z'$, $i'+z'$ filters because the dips are
faint in these wavelengths, as well as the data taken in the night of 
2014 April 3. Our target was setting with airmass $> 2$ during our 
observation on 2014 April 3, and the part of light curve with dips is 
heavily affected by noise. 
Figure~\ref{dips2014} shows 18 light curves and the flux of each 
light curve has been rescaled to 0 and 1, as in Figure~\ref{7jan}. 
For the nights where we used a short exposure time ($<5$\,s), 
the data points are binned to show more clearly the profile of the dips.  

The analysis of the dips in J1021+1744 is quite challenging, due to the 
fact that they were evolving rapidly in time 
(as seen in Figure~\ref{dips2014}) and in shape, from one simple 
structure into a complex one or vice versa.  
We decided to mark the well-visible dips, but only those 
which can be seen in almost every light curve. We used a numbering 
system from 1 to 5 based on their position in orbital phase. 
The number can be followed by letters a, b, and c for a dip which 
is split into a few smaller dips (in the case of dip 1).

Dips 1 and 3 are always present in our 2014 data. 
We marked dip 1 as `$1_{\small \textrm b}$' for the first seven 
light curves, and then assign the letters `a' and `b' after it split into 
two narrow dips. Dip 2 was marked for the first time on 2014 Jan 15, 
although, it is possible that this dip was already present in the light 
curves 
prior to this date. However, the long exposure used for the first few 
light curves does not allow us to resolve this dip.
Dips 4 and 5 were 
not present at all at the beginning of our observations. They first 
emerged on 2014 January 28 and then disappear and reappear throughout 
2014.  There were two occasions where another dip appeared 
between dips 1 and 2, which was on March 30 and April 1. This dip is 
marked as `$1_{\small \textrm c}$'. Dip $1_{\small \textrm c}$ is a fine 
example to show the swift evolution of the dips. On the night of 2014 
April 1, we observed J1021+1744 uninterruptedly for 5 hours, following two 
eclipses in orbital cycle 596 and 597. During this observation, we 
witnessed the appearance of dip 1c in cycle 597, blocking half of the 
total light from the binary for more than two minutes. Such a dip was not 
recorded in the light curve of cycle 596.

We followed the same procedure to mark the dips in our 2015 data 
(Figure~\ref{dips2015}). It is obvious that the dips which were 
present in our 2015 light curves are different from those that appeared
in our 2014 data. We obtained our first data of the second observing 
season on the night of 2014 December 22. The dip was absent from 
this light curve. 
A small dip seems to be visible at phase 1.07 on 2015 January 1 and 
January 12. 
However, we are not certain of this because it lies far (in phase) 
from the previous known dips in this system.
It is also only marginally significant given the errors and the 
fluctuations in the light curves. 
Our WHT+ULTRACAM data, which was taken four nights later, confirmed 
the presence of this small dip. This light curve also revealed 
a second shallow dip at phase $\sim$1.15 and possibly even a third dip at 
phase $\sim$1.22. Our further TNT observations 
show that only dip 2 which remains present in our subsequent 2015 
data.     

\section{Discussion}

\begin{figure*}
 \centering
 \includegraphics[scale=0.9]{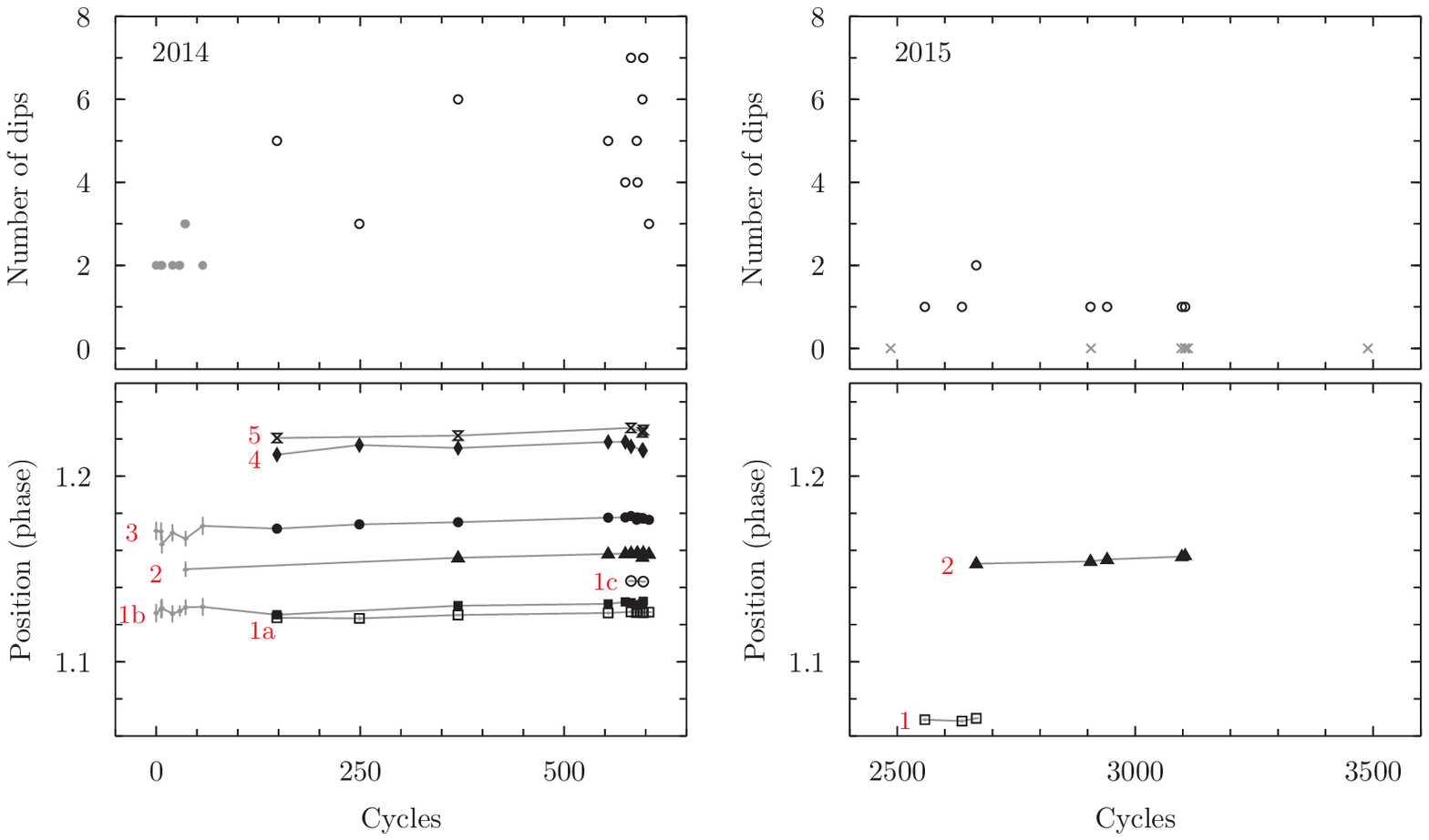}
 \caption{Top panels: The total number of dips seen in each light curve 
 in 2014 (left) and 2015 (right). The grey dots on the left panel 
 represent the number of dips in the light curves taken with integration 
 times longer than 30 seconds. The grey crosses on the right panel mark 
 the orbital cycles where the dips are absent in the light curve. Bottom 
 panels: The position of each dip in orbital phase. The data with longer 
 exposures taken between Jan $7-15$ are faded out with grey error bars.}
 \label{dipsnum}
\end{figure*}

Our data set is sufficiently extended, and the number and positions
of the dips sufficiently complex, that it is difficult to provide
a detailed discussion of each feature. However, we can discuss
in broad terms at least the time scale of the phenomenon and
the time evolution of the dips, in order to infer some conclusions.

We counted the total number of dips in each light curve based on the 
markings given in Figure~\ref{dips2014} and \ref{dips2015}, where each 
mark represent one dip. In some light curves we assigned `${}_{\small 
\textrm a}1_{\small \textrm b}$' and this mark is counted as two dips.
The result is shown in Figure~\ref{dipsnum} (top panels). In 2014, 
we found that there are dips in every light curve, starting with two 
and increasing to about five or more by 2014 April. The actual number 
varies from day to day, even from one cycle to another. 
The light curves with the highest number of dips are the $g'$-band 
light curves on 2014 March 30 and April 1, where we counted 7 dips 
in each dataset. For our 2015 observations, on the other hand,
only the ULTRACAM data showed two, possibly three, dips. We marked only 
dip 1 and 2 in the ULTRACAM data because the third dip was very weak and 
was not seen in the other light curves (possibly due to less resolution in 
the TNT data). Five 
other datasets show only one small dip, and the rest (marked with grey 
crosses) have no dip feature. 
Given the large gap between the two data sets (over 8 months without
any data), it is not possible to say whether the dips seen in 2014
evolved into those seen in 2015. 
At the very least, we can state that the dips had a lifetime of
about three months: they were well developed when we first detected
them in 2014 January, and showed no signs of abating by April.
If, on the other hand, we assume that the dips were indeed present
and evolving between April and December, then their minimum lifetime
would be at least 1.5 years. A connection with the first detected
dip by P13, implying a lifetime of several years,
seems more difficult to defend.
We recall the case of QS Vir \citep{2003MNRAS.345..506O}, in which two 
deep dips were detected before the primary eclipse. However, further
observations did not detect the dips again, pointing to a short-lived
phenomenon.

In Figure~\ref{dipsnum} we also plot the position of the dips in the
2014 and 2015 data (bottom panels). To measure the positions, we 
visually inspected each dip and tried to determine the minimum of each 
feature, unless they had an asymmetric shape. In this case we used 
the position of the data points with the lowest flux. 
Our analysis shows that all dips were shifting towards later orbital 
phase, although, small variations exist in the early 2014 data. 
From this plot we can infer that the dips are not stationary 
with respect to the orbital phase of the binary. It might 
also imply that the material is slowly drifting away from the binary.
We measure a shift of 0.01 in orbital phase for dip 3, or 
almost 2 minutes in time. Despite having different characteristics
(in depth, width, and numbers), the dips from our 2015 datasets 
also show similar behaviour. 
The fact that the dips are multiple and are shifting in phase leads
to the conclusion that the material is in the form of several blobs,
which are orbiting the red dwarf but at the same time subject to
varying gravitational forces which change their relative position
from the star and among themselves.

\begin{figure*}
\centering
\begin{minipage}{.45\textwidth}
  \centering
  \includegraphics[width=1\linewidth]{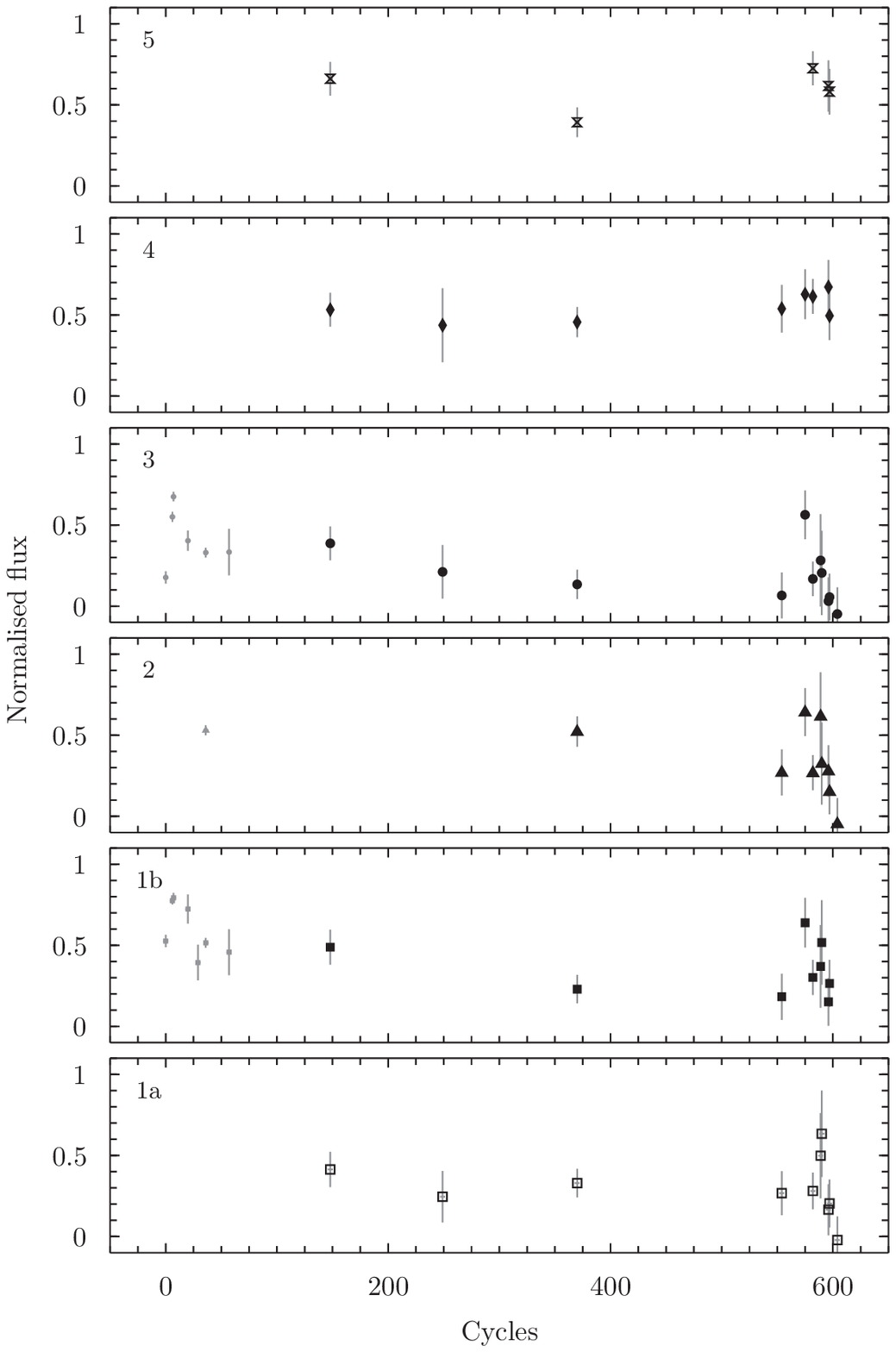}
\end{minipage}%
 \hspace{20pt}
\begin{minipage}{.45\textwidth}
  \centering
  \includegraphics[width=1\linewidth]{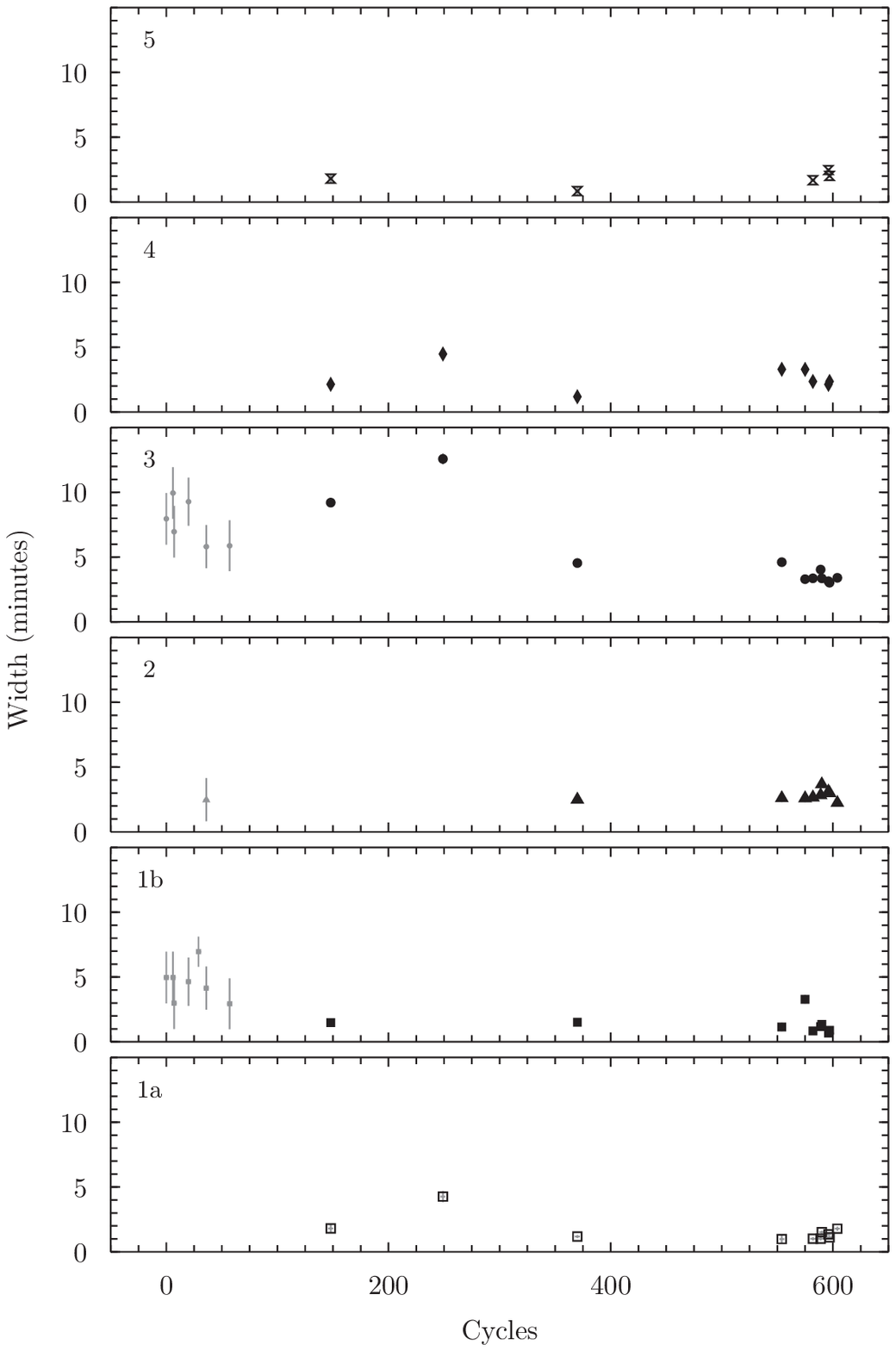}
\end{minipage}
  \caption{The plot of the flux (left) and the width (right) of the 
  dips during the 2014 observation. The flux is rescaled to 0 during 
  eclipse and 1 outside the eclipse (see text for details). In this 
  manner, a value of 0 shows that there is more light being blocked by the 
  dip. The out-of-eclipse part in the $r'$ filter is slightly 
  higher due to the contribution from the secondary star, and the dips are 
  shallower in the red filter. The width is measured at the top of the 
  dips. The grey points (see 1b, 2 and 3) are for data with larger 
  uncertainties (due to longer integration times and less resolution) 
  obtained between January $7-15$.}
  \label{dipswidth}
\end{figure*}

Our evaluation of the depth and the width of the dips are presented in 
Figure~\ref{dipswidth}. The measurement was done for the 2014 data (except 
dips 1c). We calculated the flux at the position/phase measured in 
Figure~\ref{dipsnum}. The width is measured at the level where the 
normalised flux is equal to 1. In the case of connecting dips, such as 
dips $1_\mathrm{a}$--$1_\mathrm{b}$ and dips 2$-$3, the flux values 
between the dips are often lower than 1. For this situation, we mark the 
start (or the end) of one dip at the phase with the highest flux between 
the two dips.   

The intensity plots show a similar feature where all of the dips 
are found to fluctuate on short time scale (days). This
fluctuation can be seen clearly for dips 1b and 3  
after cycle 550 in all of the dips. This short time scale variability 
is also seen in the plots of the width of the dips. These variations are 
also seen in the intensity and width plots during cycle 0$-$57. However, 
the small fluctuations are much harder to detect with the longer 
integration time, and the width is also difficult to be measured 
accurately. Hence, we faded out the data points for the first few light 
curves in 2014 January (cycle 0$-$57) as these points have larger 
uncertainties compared to the rest of the data. 
For both the intensity and the width plots, we only have three data 
points (taken in Jan 28, Feb 11, and Feb 28) between cycle 50 and cycle 
550. Therefore, we cannot tell whether there was any short time-scale 
variability during that period.

\begin{figure}
 \centering
 \includegraphics[scale=0.44,angle=-90]{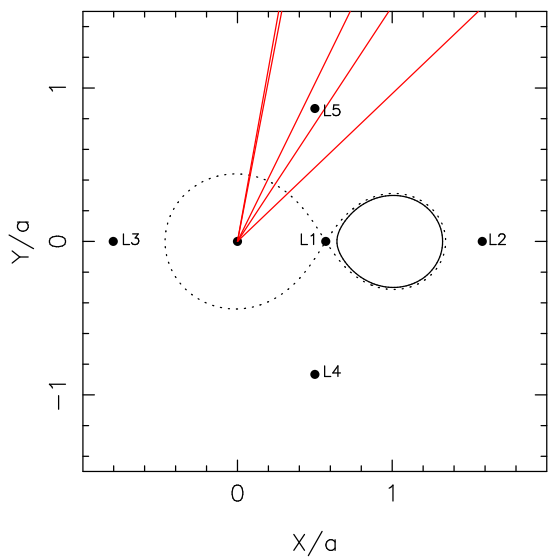}
 \caption{The Roche geometry of J1021+1744, scaled by the separation of 
 the two stars. The dotted lines indicate the Roche lobe, while L1--L5 
 mark the positions of the Lagrangian points in this binary. The red dwarf 
 star (solid black line) is seen very close to filling its Roche lobe. The 
 straight lines indicate the line of sight to the white dwarf where the 
 dips are seen on 2014 April 1. The binary rotates in clockwise direction.}
 \label{rocheplot}
\end{figure}

We note that potentially similar dips were observed before in
QS~Vir (\citealt{2003MNRAS.345..506O}) in the optical, and in V471~Tau 
(\citealt{1986ApJ...309L..27J}) in X-rays. 
However, this is the first time that such dips 
are well resolved in time and monitored over about 1.5 years
at several wavelengths. The dips in QS Vir were also detected 
spectroscopically by \cite{2011MNRAS.412.2563P}, altough the material 
was not optically thick. Therefore, the dips in QS Vir as reported by 
the authors were seen only in the lines and not in the continuum light. 
In Figure~\ref{rocheplot} we show the expected location of the dips in 
J1021+1744 as observed on 2014 April 1, where they seem to cluster 
near the L5 point. Only little force is needed to hold material at 
an equilibrium point, which might explain why we found the materials in 
J1021+1744 near the Lagrange L5 point. Opposite to the usual convention, 
the binary rotates clockwise in this figure. 

It is interesting that in QS~Vir and V471~Tau the dips are also 
reported to be in a very similar location close to the L4/L5 ``trojan"
points. \citet{1986ApJ...309L..27J} found that 
the X-ray dips in V471~Tau were seen near both L4 and L5 points, while 
the prominence material in QS~Vir is located close to its L5 point 
\citep{2011MNRAS.412.2563P}.

\begin{figure*}
 \centering
 \includegraphics[scale=1.0]{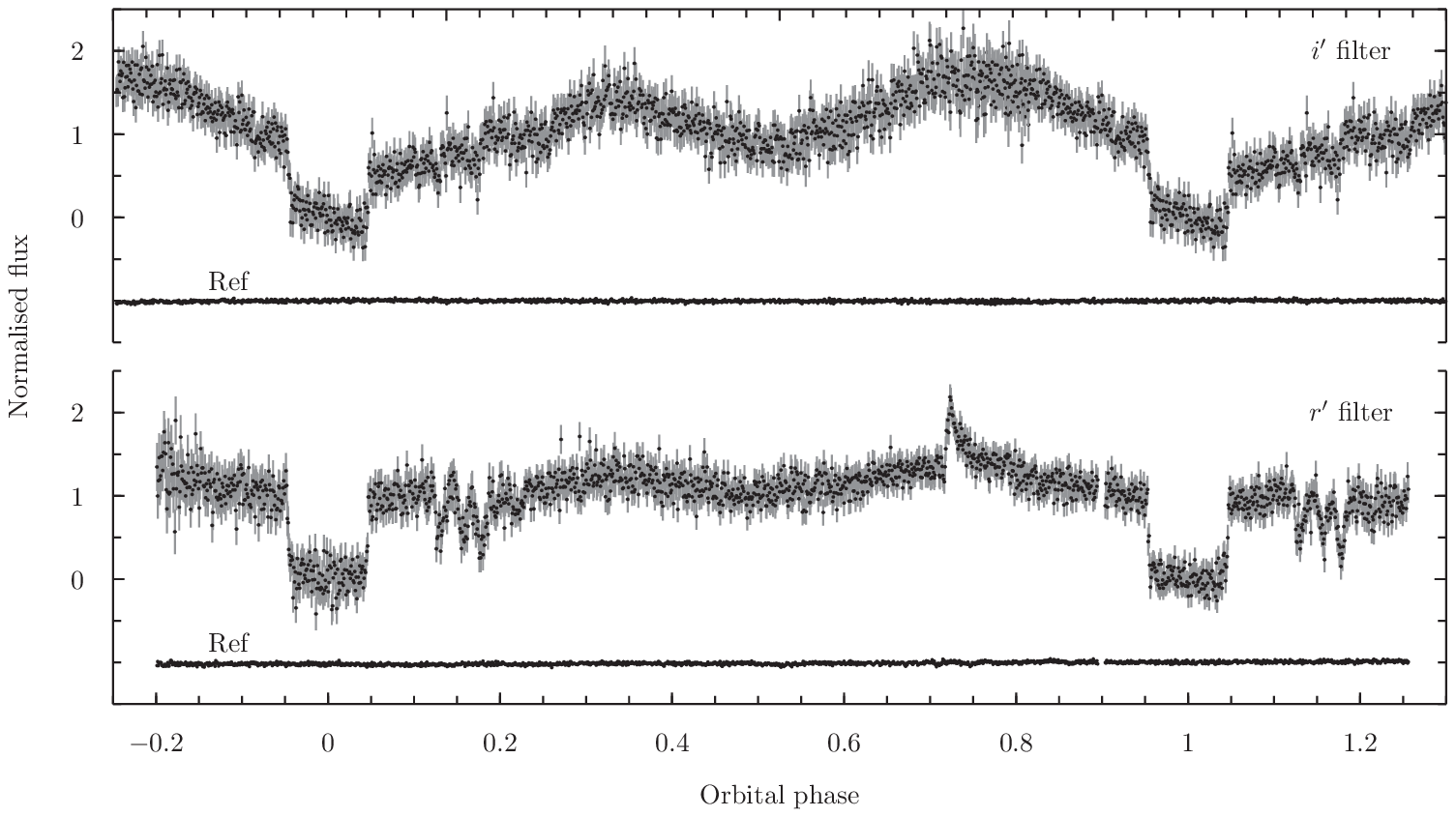}
 \caption{Full orbit light curves in $r'$ (2014 Mar 31) and $i'$ (2014 Jan 
 31) filters. The y-axis are scaled to 0 during the eclipse and 1 
 on the out-of-eclipse part (see text). The light curves of the reference 
 star are marked as ``Ref". Ellipsoidal modulation 
 is seen in both light curves, however this effect is more dominant in 
 the $i'$ filter. The $r'$-band light curve also exhibit a large flaring 
 event at phase $\sim$0.72. There is no apparent signature from the 
 low-mass star around orbital phase 0.5.}
 \label{fullorbit}
\end{figure*}

As a last remark, we report that we could, in a few cases, monitor 
J1021+1744 over a full orbit (see entries no 12, 18, and 19 in 
Table~\ref{tab1}). The data were taken in $i'$, $r'$, and $g'$, 
consecutively. Two such light curves are presented in Figure 
\ref{fullorbit}. We, unfortunately, were unable to detect the dips' 
material passing in front of the M dwarf. The secondary eclipse was also 
undetectable in any filter.
We note however that on March 31 we observed a significant brightening, 
visible around phase 0.72. The total intensity appeared to double 
and fade back within a few minutes.
This is interpreted as a flare from the red dwarf, pointing to
significant chromospheric activity. Whether such active behaviour is
partly responsible for mass ejections, which could funnel material to
the observed dip positions, is an interesting possibility.

\section{Conclusion}
We have detected the signature of dips in the light curve of
the detached, eclipsing white dwarf/M dwarf binary SDSS J1021+1744.
Although potentially similar dips were seen before in a few other stars,
such as in QS Vir (\citealt{2003MNRAS.345..506O}) and in V471 Tau 
(\citealt{1986ApJ...309L..27J}), this is the first time that such dips 
are well resolved in time and monitored over about 1.5 years
in various filters across the whole visible spectrum. 
The dips are at locations which appear consistent with being close
to the L5 point. 

Our observations show that the dips are visible over hundreds
of orbits, from a minimum of 3 months, possibly up to 4$-$6 months and 
even up to 3 years. They also clearly reveal a complex dip structure,
with their number, depth, and shape changing in time and as
a function of wavelength. The dip lifetimes are 3$-$100 times longer 
lived  than prominences on the Sun. On the other hand, the obscuration is 
probably also comparably larger, suggesting significant mass and density 
of the blobs. It is noteworthy that the dips have depths of as much as 
30\% of the total light in the $u'$ and $g'$ band, showing that the 
material absorbs continuum and not just emission lines as in the case of 
the Sun.

The origin of these dips, which we speculate is in the form
of blobs of gas or very extended prominences from the red dwarf star, 
is a new phenomenon to be reckoned with in models of PCEBs.
Future monitoring of this binary, and other similar systems, is of
crucial importance to understand the frequency of these occurrences
and to learn more about their nature.

We have also provided a new ephemeris for the binary system, significantly
improved over that of P13 thanks to a much longer time span. At the 
accuracy level of our data, where the majority of the data in the O$-$C 
diagram are scattered within $\pm10$ seconds from the zero value, we find 
no evidence of changes in the primary mid-eclipse times.


\section*{Acknowledgments}

We thank the anonymous referee for valuable comments and suggestions which 
helped improved this paper. We also thank Boris Gaensicke for useful 
discussions. PI acknowledges the support from NRC-Thailand 
and a Royal Society International Exchange. TRM acknowledges the support 
of the Royal Society International Exchange Grant and of the Science and 
Technology Facilities Council under grant number ST/L000733.  
NS and KC would like to thank Suranaree University of Technology and the
Office of Higher Education Commission for the partial support under the 
NRU project. SGP acknowledges financial support from FONDECYT in the form 
of grant number 3140585.  
This work has made use of data obtained at the Thai National Observatory 
on Doi Inthanon, operated by NARIT, and the WHT on La Palma.




\bibliographystyle{mnras}
\bibliography{j1021_bib} 








\bsp	
\label{lastpage}
\end{document}